\documentclass[aip,showpacs,reprint,jcp]{revtex4-1}
\pdfoutput=1

\usepackage{graphicx}

\begin{document}

\title{Coarse-grained simulations of DNA overstretching}
\author{Flavio Romano}
\author{Debayan Chakraborty}
\thanks{Present address: Department of Chemistry, University of Cambridge,
Lensfield Road, Cambridge CB2 1EW, United Kingdom}
\author{Jonathan P.~K.~Doye}
\thanks{Author for correspondence}
\affiliation{Physical and Theoretical Chemistry Laboratory, 
  Department of Chemistry, University of Oxford, South Parks Road, 
  Oxford, OX1 3QZ, United Kingdom}
\author{Thomas E.~Ouldridge}
\author{Ard A.~Louis}
\affiliation{Rudolf Peierls Centre for Theoretical Physics, 
 University of Oxford, 1 Keble Road, Oxford, OX1 3NP, United Kingdom}

\date{\today}

\begin{abstract}
We use a recently developed coarse-grained model to simulate the overstretching
of duplex DNA. 
Overstretching at 23$^\circ$C occurs at
74$\,$pN in the model, about 6--7\,pN higher than the experimental value at equivalent
salt conditions.  Furthermore, the model reproduces the temperature dependence
of the overstretching force well.  The mechanism of overstretching is always
force-induced melting by unpeeling from the free ends. That we never see S-DNA
(overstretched duplex DNA), even though there is clear experimental evidence
for this mode of overstretching under certain conditions, suggests that S-DNA
is not simply an unstacked but hydrogen-bonded duplex, but instead probably has
a more exotic structure. 
\end{abstract}

\maketitle

\section{Introduction}

DNA {\em in vivo} is not just a passive molecular bearer of information, but is
an active molecule that is able to respond structurally 
to cellular signals, be it through changes in
solutions conditions, protein binding or the action of molecular machines.
Some of these changes are mediated by very specific responses to the
biochemical details of the protein binding, say, but in many instances this
control is achieved through the mechanical response of DNA, e.g.\
enzymes that apply structural control by adjusting the supercoiling of
DNA.
For these reasons, there has been much interest in the fundamental mechanical
properties of DNA,\cite{Bustamante03,Peters10,Vafabakhsh12} particularly as 
there are now the means to study these properties in unprecedented detail using 
single-molecule techniques.

One particular focus has been the response of DNA to tension. At low forces the
B-DNA duplex responds elastically and is well described by the worm-like chain
model.\cite{Bustamante94} However, at higher forces, typically in the 60--70\,pN
range at room temperature (although the precise value depends on solution
conditions\cite{Williams01b,Wenner02,Zhang12}), DNA undergoes a dramatic
overstretching in which it extends by about 70\% over a few pN. Perhaps
somewhat surprisingly, even though the first detailed results on DNA
overstretching were reported in 1996,\cite{Smith96,Cluzel96} the nature of the
overstretched state is only now beginning to be fully resolved and has been a
controversial topic for much of this period.  

The two main proposals are that yielding corresponds to a transitions to a new
overstretched form of double-stranded DNA (normally termed
S-DNA),\cite{Smith96,Cluzel96} or to force-induced melting.\cite{Rouzina01,Rouzina01b,Williams02b}
Features interpreted as pointing to an S-DNA mechanism include (i) the lack of 
complete strand separation after overstretching, as might be expected for 
force-induced melting, (ii) that the force-extension curves after 
overstretching often do not initially follow that expected for single-stranded DNA 
(ssDNA), 
(iii) that further transitions can sometimes be seen at higher force (that are taken
to correspond to force-induced melting and strand dissociation) and (iv) the 
reversibility of the transition under some conditions. 
By contrast, the dependence of the overstretching force $F_\mathrm{over}$
on temperature\cite{Williams01,Zhang12} and solution conditions 
(e.g.\ pH\cite{Williams01b} and salt concentration\cite{Wenner02,Zhang12})
fits well with the force-induced melting picture; i.e.\ conditions which 
destabilize the duplex (e.g.\ higher temperature, lower salt) lead to a lower
overstretching force. Furthermore, the hysteresis sometimes seen has been 
interpreted as due to the slow nature of reassociation after the melting 
of a long DNA molecule. 

New impetus has come to this debate from the recent fluorescence experiments of
van Mameren {\em et al.} that clearly showed that at room temperature and at
5--150\,mM salt overstretching occurred by unpeeling from the free
ends.\cite{vanMameren09} Although this experiment was interpreted by some as
proof of the force-induced melting hypothesis (and hence that the S-DNA
hypothesis was unnecessary),\cite{Williams09,Chaurasiya10} this conclusion was
disputed by others.\cite{Whitelam10c,Whitelam10d,Krichevsky10} Furthermore,
subsequent
experiments\cite{Calderon09,Murade10,Fu10,Fu11,Paik11,Bianco11,Maaloum11,Gross11,Zhang12,Bosaeus12}
have shown this interpretation to be too simplistic, and provide a basis for a
more balanced position. In particular, these experiments have shown that the
two modes of overstretching can occur depending on conditions, with S-DNA being
favoured at low temperature, high salt, high G-C content and short time scales.
One of the complications is that at room temperature the
two mechanisms can compete. So for lower salt concentrations unpeeling is
dominant, as in the experiments of van Mamaren {\em et al.}\cite{vanMameren09}
and the impressive high-resolution follow-up study of Gross {\em et al.} which
resolved a saw-tooth structure in the overstretching force-extension curves
that correlated with the ease with which different parts of the sequence could
unpeel.\cite{Gross11}   However, at higher salt, an S-DNA mechanism becomes
more feasible. For example, in the elegant experiment of Paik {\em et al.} at
150\,mM salt the transition occurred with some hysteresis when free ends were
present (presumably due to unpeeling), but occurred reversibly at the same
overstretching force for a DNA construct that had no free ends and was still
torsionally unconstrained (presumably by an S-DNA mechanism).

There has also been a lot of work using theory and simulation that has 
aimed to provide further insights into the nature of the overstretching
transition. The theoretical approaches usually use simple polymer models
to describe each of the different possible states
with parameters fitted to reproduce experimental behaviour.\cite{Cizeau97,Ahsan98,Storm03,Cocco04,Hanke08,Rahi08,Whitelam08b,Whitelam08c,Whitelam10c,Chakrabarti09,Marenduzzo09,Einert10,Massucci10,Nisoli11,Fiasconaro12,Manghi12} 
In particular, the work of Cocco and Marko,\cite{Cocco04} 
and Whitelam and coworkers\cite{Whitelam08b,Whitelam08c,Whitelam10c} 
has provided important insights into the competition between different
processes, such as S-DNA formation versus unpeeling. Furthermore,
such approaches are particularly well-suited to the long length and time scales
that are typical of experiments on DNA stretching. However, as these models
lack a detailed molecular representation of the underlying phenomena,
the questions these approaches can address is inevitably limited.

At the other extreme, one has simulations of DNA stretching using 
fully-atomistic models.\cite{Lebrun96,Harris05,Lohikoski05,Wereszczynski06,Luan08,Li09c,Santosh09,Rezac10,Balaeff11,Santosh11,Wolter11}
These have the potential to provide answers to detailed 
structural questions that are inaccessible to experiment. 
However, firstly, the computational cost of such simulations means that the 
pulling rates are often extremely fast, making it likely that much of 
what is 
observed is not at equilibrium. Secondly, it is not clear how well the atomistic
potentials will perform under the more extreme conditions associated with 
overstretching. Thirdly, 
in some of the simulations, constraints or boundary conditions are used that 
restrict the range of possible behaviour, e.g.\ torsional constraints or the absence of free ends. For these reasons, it is perhaps not
surprising that these simulations only observe overstretching at forces
well above the experimental values (all above 100\, pN, some substantially 
more) and that the transitions are much broader.
Furthermore, although they have the potential to identify 
the structure of S-DNA, no consensus has been reached.

In the middle are simulations using coarse-grained models that have a simplifed 
representation of DNA and so allow longer time-scale and length-scale 
processes to be more easily accessible. These can be divided into two types. 
Firstly, simple models that attempt to obtain qualitative insights into 
the stretching behaviour that stem from capturing some of the basic physical 
ingredients of DNA.\cite{Marenduzzo09,Marenduzzo10,Singh10} Secondly, there are 
coarse-grained DNA models that are fitted in order to provide a 
quantitatively accurate description of DNA's 
behaviour.\cite{dePablo11,Lankas11,Doye12} For some of these models, the 
stretching behaviour has been simulated.\cite{Niewieczerzal09,Savin11b,Kocsis12,Hsu12}
However, the observed overstretching forces are also typically much higher 
than experiment (e.g. 320\,pN in Ref.\ \onlinecite{Savin11b}, 500-1100\,pN in Ref.\ \onlinecite{Kocsis12} and 400-550\,pN in Ref.\ \onlinecite{Hsu12}).

Here, we simulate DNA overstretching for our recently developed coarse-grained
DNA potential, which provides a quantitative description of many 
of the structural, thermodynamic and mechanical properties of 
DNA.\cite{Ouldridge10,Ouldridge11} One aim of the work is to provide an 
independent test of the robustness of the model by applying it to phenomena 
to which it has not been fitted. A second aim is to provide fundamental 
insights into the nature of the overstretching transition, both by visualizing 
some of the mechanistic aspects of the transitions, and by deducing from the 
successes and failures of the model which ingredients are needed to capture 
different aspects of the transition.
We proceed as follows. In the Methods section, 
we describe our DNA potential, the simulation techniques
used and some of the theory of the thermodynamics of pulling. 
In the Results section, we focus in detail on the room temperature 
overstretching behaviour, before more briefly examining
the temperature dependence of overstretching, including looking for the
re-entrant behaviour that has been predicted to occur near melting.\cite{Rouzina01,Williams01,Hanke08,Rahi08,Einert10,Marenduzzo09,Marenduzzo10}

\section{Methods}

\begin{figure}
\includegraphics[width=8.5cm]{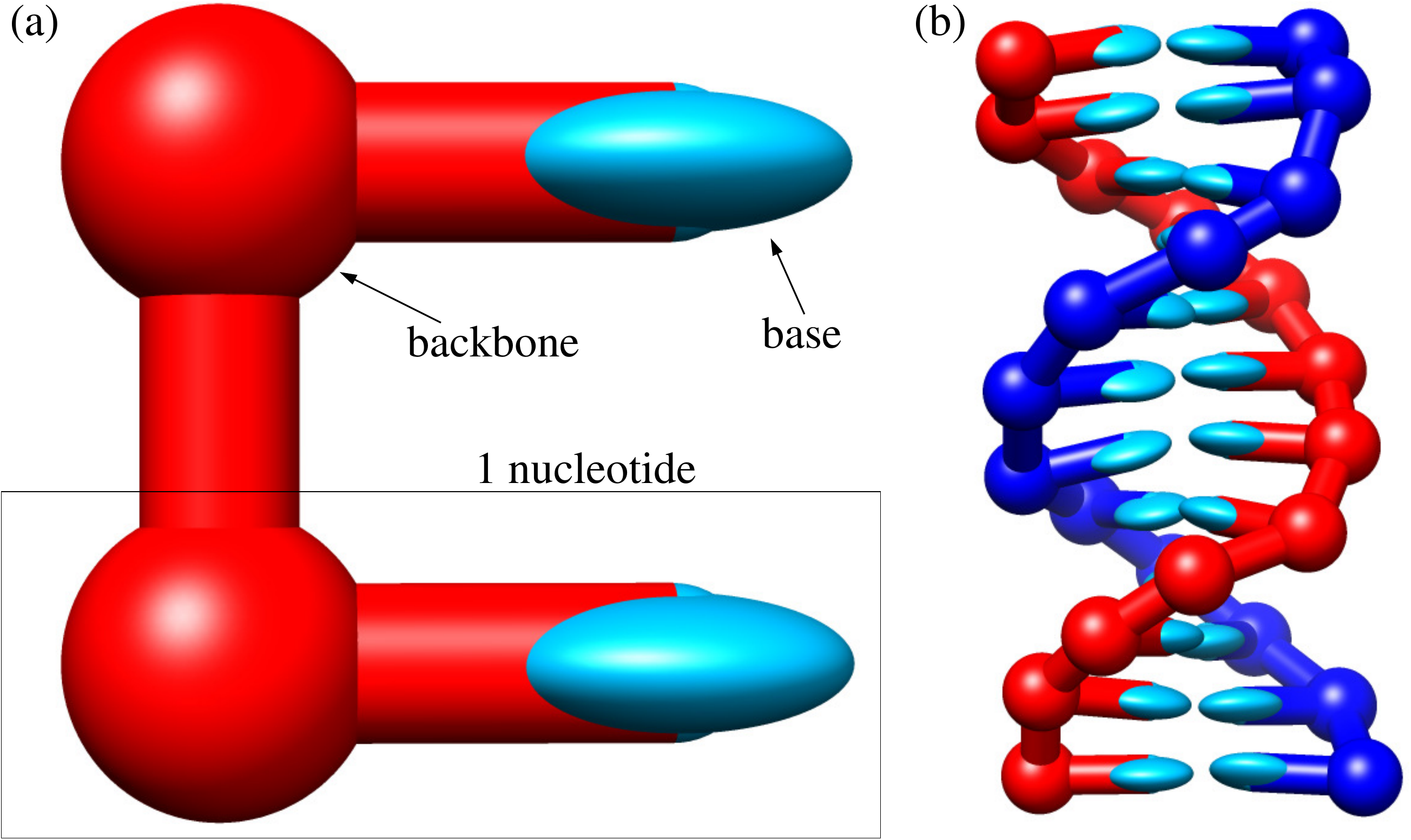}
\caption{\label{fig:CGmodel}
(a) Two nucleotides represented by our model, showing the rigid nucleotide
unit and the backbone and base regions. (b) A short DNA duplex as 
represented by our model.
}
\end{figure}

\subsection{DNA model}

We use the coarse-grained model of DNA recently developed in our 
group.\cite{Ouldridge10,Ouldridge11} 
In particular, the model has been designed to provide a good description of
the structural, thermodynamic and mechanical properties of both double-
and single-stranded DNA, all features that are very important if we are to be
able to describe DNA overstretching. 
In addition, the wide range of applications for which the model has been 
successfully used give us further confidence in its robustness. These
applications so far include DNA nanotweezers,\cite{Ouldridge10} 
kissing hairpins,\cite{Romano12b} DNA walkers,\cite{Ouldridge11b,Ouldridge13,Sulc13} 
the formation of cruciform DNA\cite{Matek12} and the nematic transition
of dense solutions of short duplexes;\cite{deMichele12} furthermore, 
investigations into hybridization, strand-exchange reactions, plectonemes, 
DNA origami and phase transitions in solutions of DNA junctions are ongoing.

The model describes each nucleotide as a rigid object (Fig.\
\ref{fig:CGmodel}(a)) that has interactions corresponding to backbone
connectivity, excluded volume, stacking, hydrogen-bonding between complementary
bases and cross-stacking. Each nucleotide has three collinear interaction sites
(corresponding to the centres of the backbone, stacking and hydrogen-bonding
interactions) and a vector perpendicular to this axis to capture the planarity
of the bases. We would like to emphasise that the attractive interactions in
the model are not isotropic, but depend on the relative orientation of the
nucleotides. It is these angular modulations that ensure that the DNA helix is
right-handed and pairs in an anti-parallel manner. The full form of the
potential has been given in Refs.\ \onlinecite{Ouldridge11,Ouldridge11b}.  A
simulation code incorporating the potential is available to
download.\cite{oxDNA}

We should note a number of simplifying features of the model that are relevant
to the current study. 
Firstly, we do not attempt to model the electrostatics explicitly, 
but instead have fitted the potential parameters for a particular salt 
concentration, namely 500\,mM. At this concentration the Debye screening length
is short-ranged and the effects of the electrostatics are included
in the excluded-volume interactions. 
Fortunately, this salt concentration is also one
of those commonly used in experiments on overstretching. 
However, as a consequence of this simplification we cannot of course 
examine how overstretching depends on salt concentration.
At room temperature and 500\,mM salt, experiments indicate that 
both the force-induced melting and S-DNA modes of overstretching 
compete.\cite{Fu10,Fu11,Zhang12} 

A second of the model's simplifications is that it ignores sequence
dependence in the interactions except for the Watson-Crick nature of 
the base pairing (i.e.\ the hydrogen bonding term in the potential only 
occurs for complementary bases). 
Therefore, each base pair has the same
average interaction strength irrespective of its identity and its neighbours. 
This ``average-base'' approximation can be an advantage when one is interested
in the generic behaviour of DNA, as for the most part we are here, but
of course it excludes us from, for example, 
examining how the overstretching depends on G-C content.
A sequence-dependent version of the model has very recently been 
developed.\cite{Sulc12}

Thirdly, the double helix in our model (Fig.\ \ref{fig:CGmodel}(b)) is
symmetrical with the helical grooves being of the same size. This
simplification, however, is likely to be relatively unimportant for the current
study.

\subsection{Pulling Schemes}
\begin{figure}
\includegraphics[width=8.5cm]{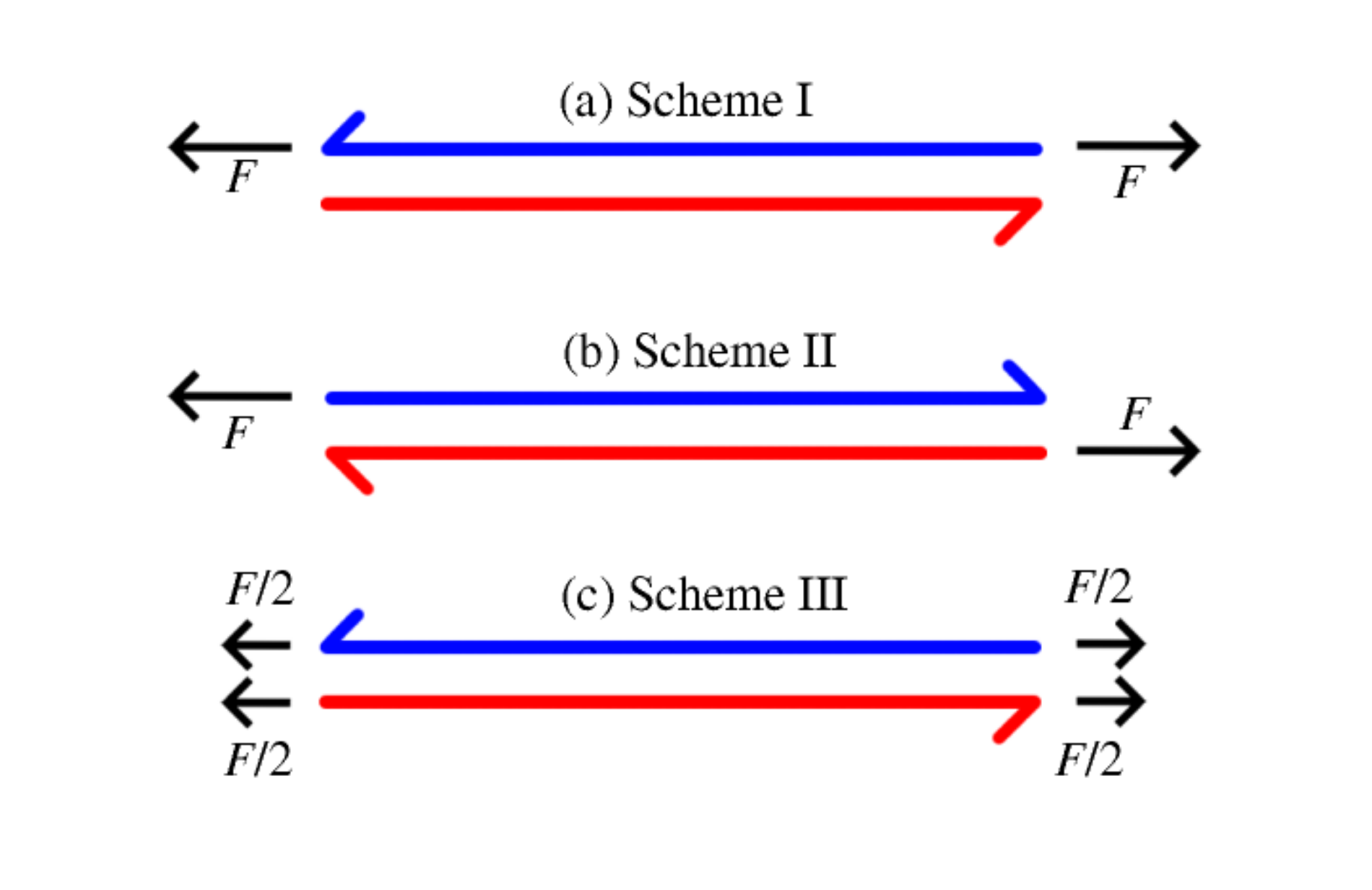}
\caption{\label{fig:schematics}
Schematic representation of the three pulling schemes used. 
The arrows on the DNA are in the $3'$-$5'$ direction.
}
\end{figure}

The three ways of pulling DNA that we will consider are shown in Fig.\
\ref{fig:schematics}. Our main focus will be on scheme I, in which a force is
imposed on both ends of the same strand, and the free ends allow the duplex to
overstretch by unpeeling. Scheme II is similar in that a force is exerted on 
only one of the two strands at each end of the duplex, thus again allowing unpeeling. 
However, as the forces are applied to different strands, dissociation will be 
completely irreversible when pulled at constant force.
In our simulations, we pull on both 3$^\prime$ ends of each strand, but given the
symmetric nature of the helix in our model and the simplified representation of
the nucleotides, we do not expect any differences for pulling both 5$^\prime$ ends.
In experiments differences between 5$^\prime$-5$^\prime$ and 3$^\prime$-3$^\prime$ pulling 
are only found at forces well above that for overstretching.\cite{Danilowicz09}

Finally, in scheme III, as both ends of both strands are pulled, force-induced
melting by unpeeling is suppressed. Therefore, if the DNA is to melt, it would have
to be by bubble formation. Note the system in scheme III is not torsionally
constrained, i.e.\ the ends are free to rotate.

\subsection{Thermodynamics of pulling}
\label{sect:thermo}

Before we consider our pulling simulations, it is important to understand the
effect of imposing a force on the thermodynamics of our 
systems.\cite{Rouzina01,Cocco04}
The force provides an additional source of work being done on the system.
Assuming the force $F$ acts along the $z$ direction, this gives 
an additional term $Fdz$ to the change in internal energy and hence to 
the change in Helmholtz free energy $dA$, and $A$ is therefore a function of $z$.
However, it will usually be more convenient to consider a system under 
constant force, and so we introduce the free energy $A'=A- Fz$ to
describe such a constant-force ensemble. 
It follows that the change in free energy on applying a force is 
\begin{equation}
A'(F)-A'(F=0)=-\int_0^F z(F)dF.
\label{eq:A_F}
\end{equation}
Therefore, the force-extension curves of dsDNA and ssDNA determine the
relative stabilization of these two forms by force.
Furthermore, assuming overstretching occurs by force-induced melting, one
just needs a correct description of the zero-force thermodynamics and
the force-extension curves to correctly predict overstretching.

Force-induced unpeeling of DNA occurs at the force at which the free energy per
base pair of the ssDNA and dsDNA states are equal, leading to coexistence of
the two forms. In other words, at this overstretching force $F_\mathrm{over}(T)$,  the
average free energy change for a duplex to unpeel by one more base pair is
zero. We wish to emphasise that overstretching, in the force-induced melting
picture, is not determined by an equilibrium between dsDNA and fully-dissociated
ssDNA, but rather by when the duplex becomes unstable with respect to
unpeeling. In this sense overstretching is distinct from melting, as the
translational entropy gain from complete dissociation of the two strands plays
no role. However, in the bulk limit of infinitely long chains, the contribution
of  translational entropy is negligible and melting and unpeeling transitions
are equivalent. In this limit, the melting temperature $T_m(F)$ can be defined 
as the temperature at which the average free energy change for the loss of one 
base pair from a duplex is zero at a given force $F$, and the curves $F_\mathrm{over}(T)$ and
$T_m(F)$ then give equivalent representations of the overstretching transition
in the force-temperature plane. However, for finite chains $T_m(F)$ will depend 
on concentration, and this equivalence will be broken.

We also note that care should be taken when applying this approach to the
different schemes in Fig.\ \ref{fig:schematics}. 
For scheme I only one of the
strands is force-bearing when single stranded. By contrast, in scheme III the
force on each single strand is $F/2$.  Finally, in scheme II, equilibrium is
not well-defined, as for any non-zero force the most stable state is two 
dissociated single strands because they can then increase their separation without limit.

\subsection{Simulation methods}

We use a mixture of Monte Carlo and Brownian dynamics to simulate DNA stretching
for our model. We use Monte Carlo to obtain the equilibrium behaviour of the
system at a particular force, and Brownian dynamics to study the stretching 
dynamics of DNA when the force is increased or decreased (linearly)
as a function of time.

To aid the equilibration in our Monte Carlo simulations, we use the virtual 
move Monte Carlo approach of Whitelam and coworkers.\cite{Whitelam07} 
This algorithm is a type of ``cluster-move'' Monte Carlo that allows 
clusters of nucleotides to be moved at each step, where the clusters that
are constructed reflect both the configuration and the proposed move.
The collective motion of nucleotides that this algorithm introduces 
allows more efficient sampling for our model compared to Monte Carlo with 
single-particle moves. 

For the dynamics simulations we use the simple Brownian thermostat introduced 
in Ref.~\onlinecite{Russo09}. The system is evolved according to Newtonian
dynamics, but every $n_{\rm Newt}$ time steps a fraction $p$ of the
particles have all three components of their velocities drawn from the
Maxwell distribution. 
On time scales longer than $p n_{\rm Newt}$, the dynamics is Brownian. 

Modifying these simulation algorithms to incorporate the forces exerted on the
terminal bases consistent with the different schemes in Fig.\ 
\ref{fig:schematics} is straightforward.
To perform Monte Carlo at constant force  it is necessary to include
an additional term in the energy $-\sum\limits_i F_i z_i$, where $F_i$ is the
external force imposed on nucleotide $i$. In Monte Carlo, this gives rise to an
extra term in the Boltzmann factor in the standard Metropolis acceptance
criterion, namely $\exp(\beta\sum\limits_i F_i \Delta z_i)$. 
For the Brownian dynamics, implementing the constant force is just a
matter of adding the external forces to the forces acting on the relevant
nucleotides.

\section{Results}

\begin{figure}
\includegraphics[width=8.5cm]{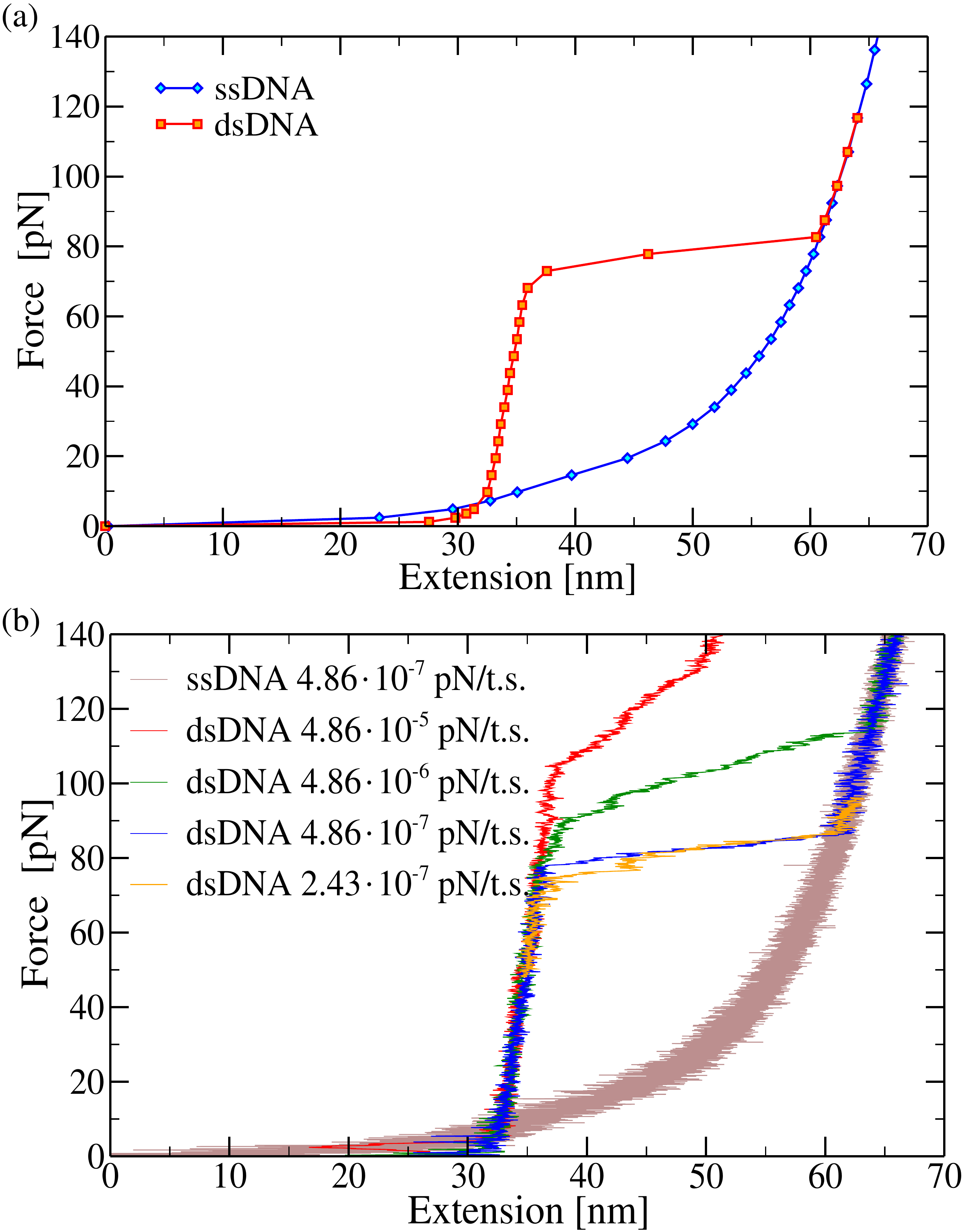}
\caption{\label{fig:Fvx_roomT}
Force-extension curves for DNA at 23$^\circ$C for pulling scheme I. 
In (a) results from a series of Monte Carlo simulations at constant force 
are presented. 
For comparison the curve for single-stranded DNA has been added.
In (b) results from dynamics simulations are presented at a number
of different pulling rates. 
The pulling rates are given in pN/time step.
These units can be converted into pN\,ns$^{-1}$ by multiplying by 
$1.17\times 10^{5}$ time steps/ns, 
but as with any coarse-grained model absolute values of time should be treated 
with caution. Using this conversion, our slowest and fastest rates correspond 
to $0.0284$\,pN\,ns$^{-1}$ and $5.69$\,pN\,ns$^{-1}$, respectively.
}
\end{figure}

In all cases, we consider a 100 base-pair duplex
with a random sequence.
We first consider pulling by scheme I in detail. Force-extension curves at
``room temperature'' (23$^\circ$C) are presented in Fig.\ \ref{fig:Fvx_roomT} 
for Monte Carlo simulations at different forces, and for dynamics
simulations at different pulling rates. This data shows a number of clear
features. 
Firstly, the MC data (which we expect to be closer to ``equilibrium'') 
shows a clear and narrow overstretching transition at a force of about 77\,pN.  
At the slowest pulling rates, the dynamics
results also show a clear overstretching plateau at a very similar force.
However, as the pulling rate increases, there is an increasing tendency to
overshoot this transition, and for overstretching to start at a higher force
and for the transition to be spread over a wider range of force, simply
because the time scale associated with the overstretching transition is no
longer fast compared to the pulling rate.
These effects of pulling rate are also relevant to the all-atom simulations as
they provide an indication of the consequences of being increasingly far from
equilibrium. Indeed, our results at faster pulling rates somewhat resemble
the force-extension curves seen in all-atom simulations.\cite{,Harris05,Lohikoski05,Wereszczynski06,Luan08,Li09c,Santosh09,Rezac10,Balaeff11,Santosh11,Wolter11}
We should also note that although our slowest pulling rates are significantly
slower than those used in all-atom simulations, they are still much faster than
in typical experiments.

\begin{figure}
\includegraphics[width=8.5cm]{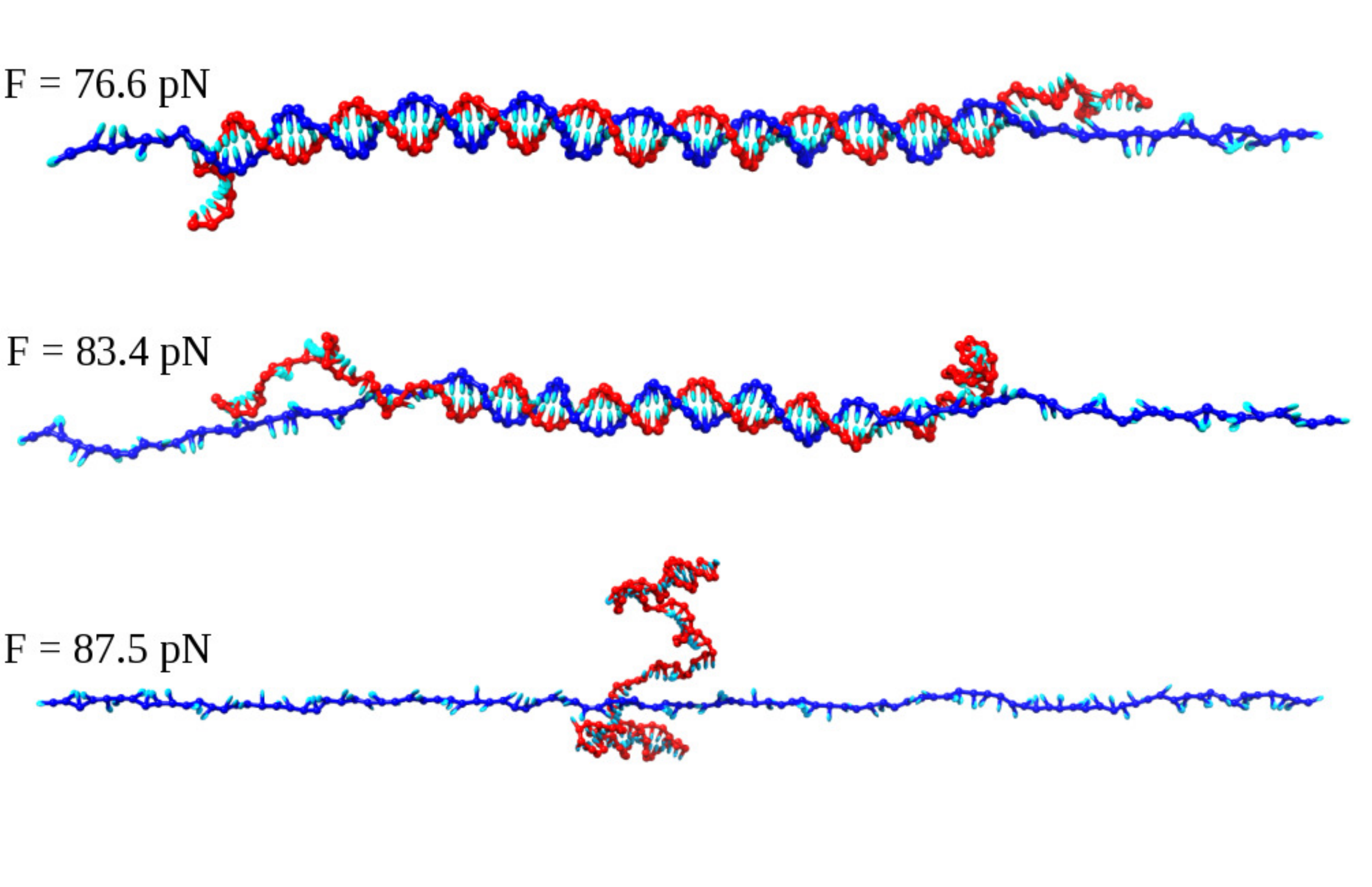}
\caption{\label{fig:pics_roomT} Snapshots of typical DNA configurations as the
system passes through the overstretching transition at 23$^\circ$C for scheme I.
The snapshots are from the dynamics run at a pulling rate of 
$2.43\times 10^{-7}$\,pN/time step.}
\end{figure}

Secondly, beyond the overstretching transition, the force-extension curves
follow that for ssDNA, thus indicating that DNA overstretching in our model is
a result of force-induced melting. This conclusion is confirmed in Fig.\
\ref{fig:pics_roomT} which depicts typical configurations as the system passes
through the overstretching transition. From these shapshots, it can be clearly
seen that the increase in extension is a result of DNA unpeeling from the free
ends. In particular, there is no sign of any S-DNA-like state. Experiments
suggest that the transition from B- to S-DNA is able to occur more rapidly than
force-induced melting.\cite{Fu10,Fu11,Whitelam10c}  Therefore, if S-DNA exists
for our model, one might expect that it is more likely to be seen in the faster
pulling simulations where the overstretching is observed at higher force.
However, in all the simulations depicted in Fig.\ \ref{fig:Fvx_roomT}(b) we
never found any evidence of an S-DNA-like state.  Since the mechanism of
overstretching for our model is always unpeeling, we should note that we expect
the dynamical behaviour of our model to depend on chain length with slower
pulling rates needed for longer molecules if overshooting of the transition is
not to occur.

\begin{figure}
\includegraphics[width=8.5cm]{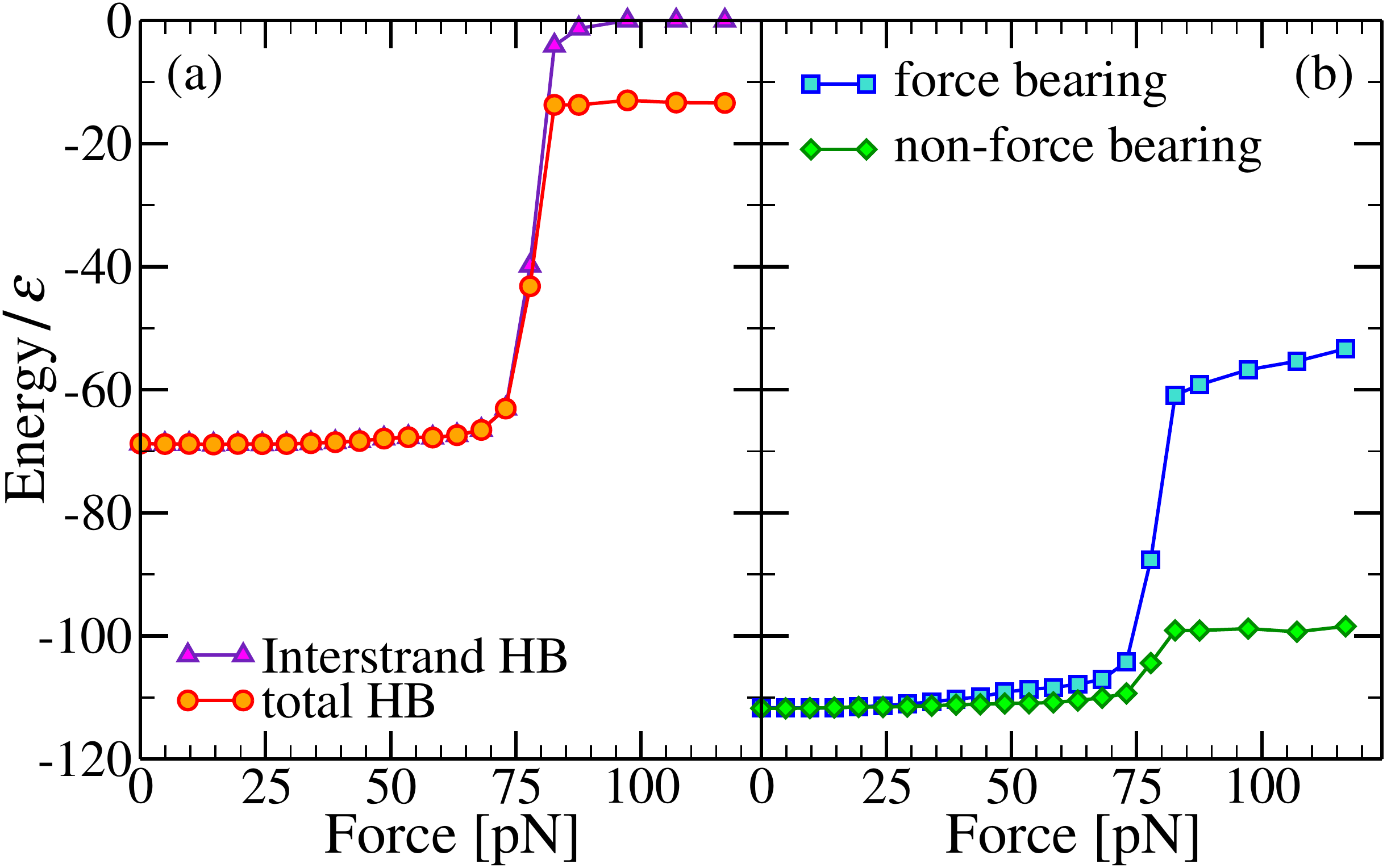}
\caption{\label{fig:V_roomT} Contributions to the potential energy
from (a) hydrogen bonding and (b) stacking as a function of force.
In (b) the contributions from the two individual strands is included. 
$k_\mathrm{B} T/\epsilon=0.1$ at $T=300\,\mathrm{K}$}
\end{figure}

There are a number of more minor features that are apparent from the
configurations in Fig.\ \ref{fig:pics_roomT}. Firstly, immediately beyond the
overstretching transition, the duplex does not fully dissociate, because for
short (e.g.\ 4 base pairs) intermolecular helices, the force-bearing strand can
still approximately align itself along the direction of the force, and so the
extension gain associated with the loss of the base pairs is lower for these
short duplexes. Therefore, a somewhat higher force is required to dissociate
these last few bases.  This effect can be seen in Fig.\ \ref{fig:V_roomT}(a)
where overstretching leads to the loss of most, but not quite all, of the energy
associated with intermolecular hydrogen bonding --- only at slightly higher
forces does this term go to zero. 

Secondly, from a careful inspection of Fig.\ \ref{fig:pics_roomT}, one can see
that the orientation of the bases in the overstretched force-bearing strand is
not random, but there seem to be short runs of stacked bases. Although
completely stacked ssDNA in our model is helical, it is possible for short
sections of 3--4 stacked bases to orient the backbone approximately along the
direction of force.\cite{Sulc12} For this reason, the force-bearing strand is
able to retain a significant fraction of its stacking interactions after
overstretching. Fig.\ \ref{fig:V_roomT}(b) indicates that just after
overstretching is complete, the stacking energy for the force-bearing strand is
57\% of that for B-DNA and 61\% of that for ssDNA at zero force (assuming the
unpeeled non-force-bearing strand is typical).  Interestingly, ssDNA bound to
oligomers of DnaA proteins adopts a similar stretched geometry with short runs
of three stacked bases.\cite{Duderstadt11}

Thirdly, it is clear from Fig.\ \ref{fig:pics_roomT} and the appearance  of
intramolecular base pairs in Fig.\ \ref{fig:V_roomT}(a) that the unpeeled strand
exhibits secondary structure, in particular hairpins. 
Even though the sequence is random, hairpins can be formed with stems
having a sufficient number of Watson-Crick base pairs (as well as some
mismatches) to be stable. 

Fig.\ \ref{fig:Fvx_roomT}(a) only provides a rough guide to the position of the
overstretching transition, because it is hard to fully equilibrate the
simulations in the vicinity of the transition. One approach to
locate $F_\mathrm{over}$, as noted in Sect.\ \ref{sect:thermo}, is to find the
force at which the average free energy for unpeeling a base is zero. However,
this approach is complicated by secondary structure formation. Firstly, it
makes thermodynamic sampling of the transition more difficult, because
secondary structure in the unpeeled state leads to free energy barriers between
different states. For example, it might be that for intermolecular base pairs
to form a hairpin must first open, or as the double helix unpeels the identity
of the most stable hairpin in the unpeeled strand changes. Secondly, it
obscures some of the basic physics of the overstretching transition, as
hairpins will stabilize the unpeeled state, but the extent to which this occurs
(and hence the overstretching force) will be sequence dependent (even for our
``average base'' model where all the interactions are assumed to be of the same
strength).

Therefore, in our simulations to locate the overstretching force, 
we turn off intramolecular base pairing\cite{nocomplement}
and, but less importantly, non-native intermolecular base pairing.\cite{repetitive} 
As a consequence, the free energy change for unpeeling every base pair is 
now zero (not just on average) at the overstretching transition of our model
and so we only need to sample a local section of the free energy landscape, 
rather than that for the whole transition.
However, this approximation will lead to an overestimation of the overstetching
force, because of the stabilization of the unpeeled state by secondary 
structure. Later, we will estimate the magnitude of the resulting overestimation.

\begin{figure}
\includegraphics[width=8.4cm]{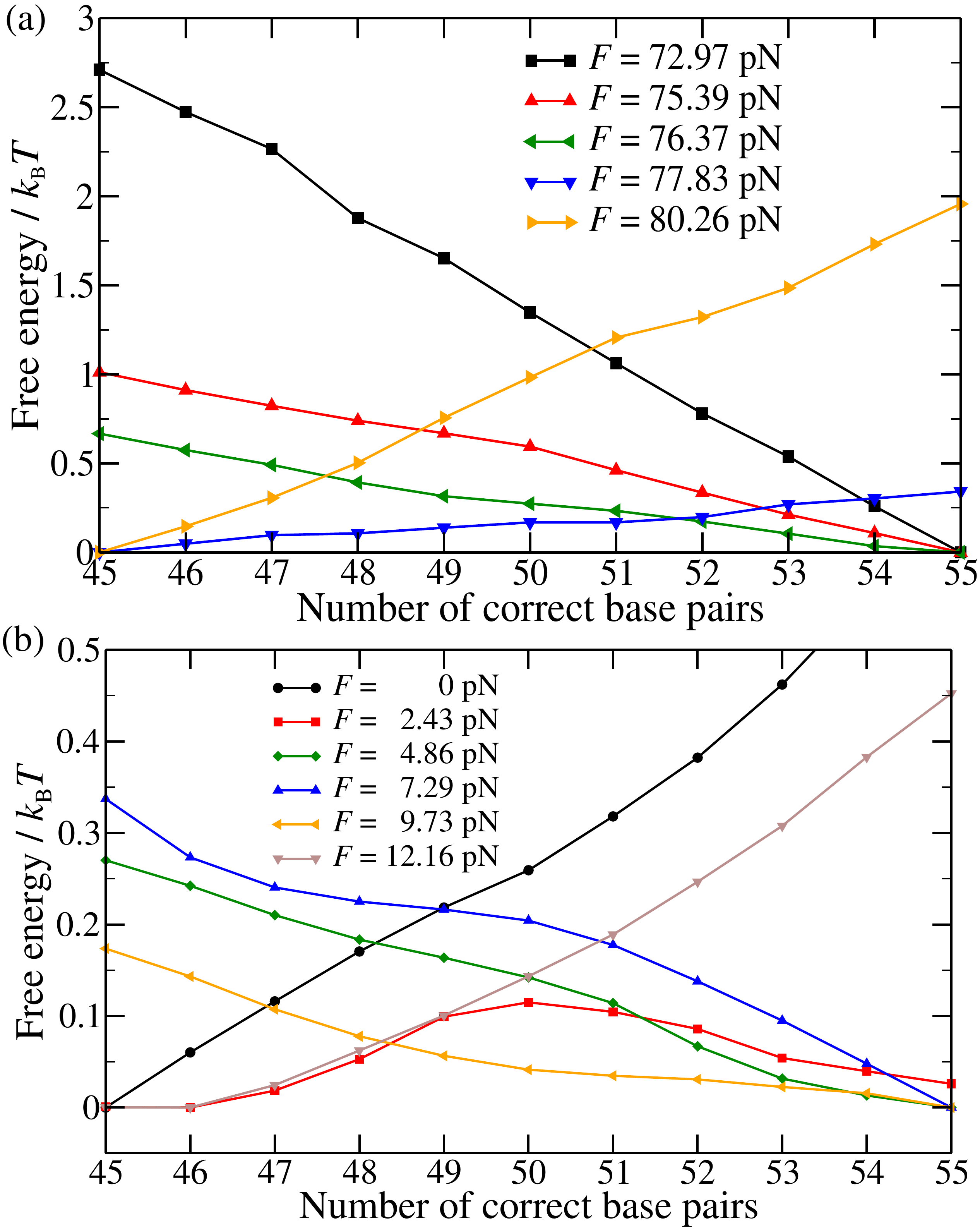}
\caption{\label{fig:FEL} Free energy profiles as a function of the
number of base pairs for different forces at (a) room temperature 
($T=23^\circ$C) and (b) $T=94^\circ$C. The latter is just above the zero-force 
bulk melting temperature and exhibits reentrance. Note the relative flatness of 
the free energy profiles at this temperature.}
\end{figure}

In Fig.\ \ref{fig:FEL}(a) we show the free energy landscape for a 10-base
section of the 100-base-pair duplex for a number of forces close to
$F_\mathrm{over}$.  In real DNA the free energy landscape for unpeeling will
have a saw-tooth structure due to the sequence-dependence of the thermodynamics
of base-pairing (e.g.\ G-C base pairs are stronger than A-T base pairs), and
the overstretching force will be where the average free-energy change for
forming a base pair is zero.  However, because of the ``average-base'' nature
of our model, 
the free energy is roughly linear in the number of base pairs in the duplex. 
When the slope is positive, the fully associated duplex is most stable, whereas
when the slope is negative the fully 
unpeeled state is most stable.  Thus, by estimating the force for which the
slope is zero, we obtain a value of $F_\mathrm{over}\approx 77$\,pN for our
model at 23$^\circ$C when intramolecular base-pairing is turned off.

On the longer time scales in experiment, when unpeeling occurs at equilibrium,
the competing states are dsDNA and ssDNA in which the free strand can form
secondary structure.  Therefore, it would be particularly useful for the
comparison with experiment if we could estimate the error in the overstretching
force that results from our constraint preventing secondary structure in the
unpeeled state. The stabilization of ssDNA by secondary structure
can be computed from the difference in the ssDNA force-extensions curves when
intramolecular base-pairing is allowed or forbidden using Eq.\ \ref{eq:A_F},
and hence we can obtain an estimate of the correction to the overstretching
force. 

\begin{figure}
\includegraphics[width=8.4cm]{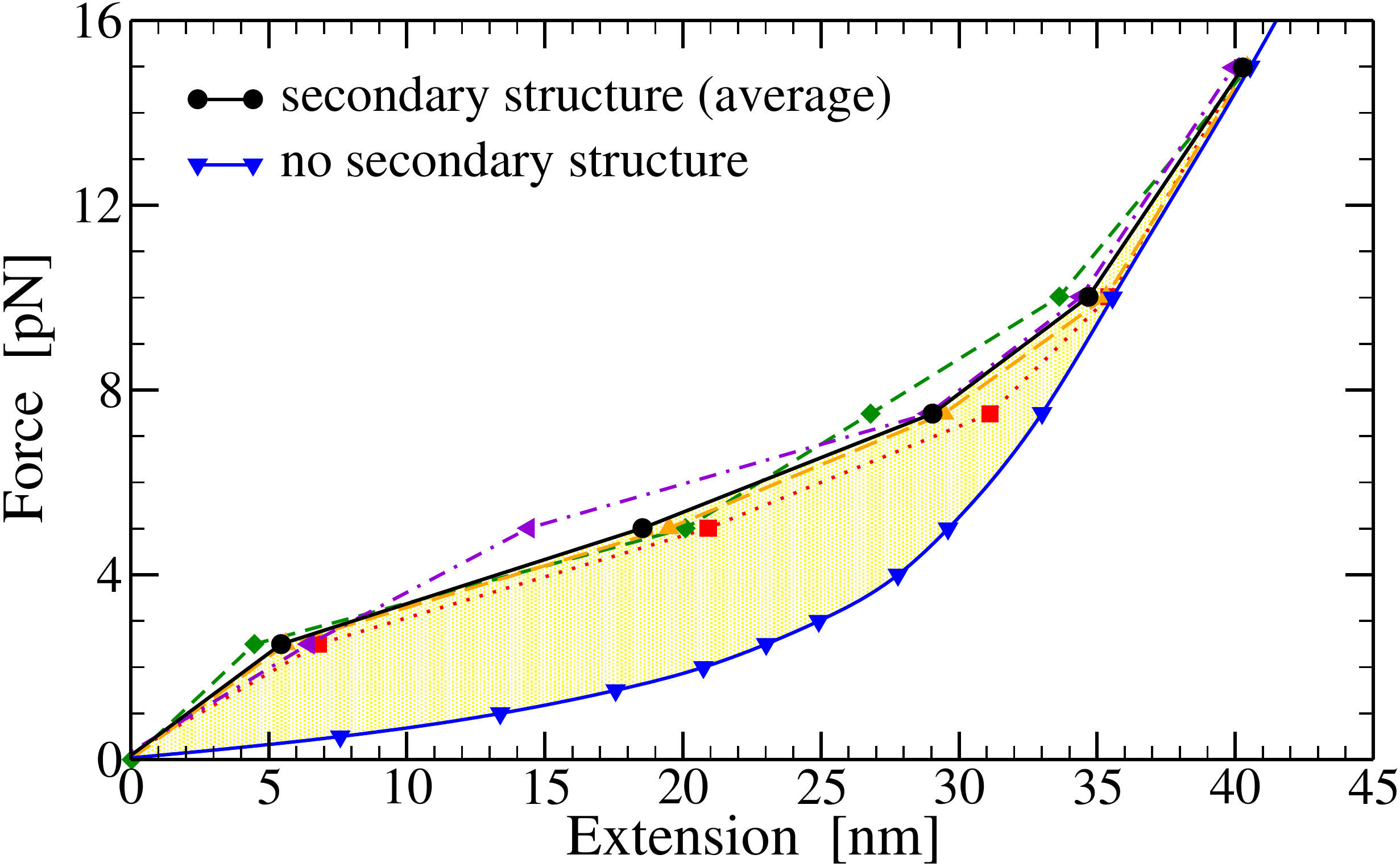}
\caption{\label{fig:Fx_ssDNA} Force-extension curves for ssDNA in the
presence or absence of secondary structure. As the effects of secondary
structure depend on sequence, the results for four different 
random sequences (as well as their average) are depicted. The area of the 
shaded region corresponds to the average free energy of stabilization of 
ssDNA by secondary structure.}
\end{figure}

Although computing the force-extension curve of ssDNA with secondary structure
is less difficult than simulating overstretching in the presence of secondary
structure, it is nonetheless challenging, because of the potentially large
free-energy barriers between different states. We therefore use parallel
tempering, in which exchanges are attempted between simulations at
different temperatures,\cite{Earl05} to help equilibration. Furthermore, as the
stabilization will be sequence-dependent, one would also want to average the
correction over different sequences when comparing to experiments for long
DNAs. We therefore performed our simulations for five different sequences.
However, for the sequence that formed the strongest secondary structure, we
were unable to achieve equilibrium. The force-extension curves of the four
other sequences are depicted in Fig.\ \ref{fig:Fx_ssDNA} (along with their
average) and compared to that for ssDNA with no secondary structure. In
particular, the secondary structure makes the ssDNA less extensible at low
force. However, as the force increases the secondary structure becomes
destabilized such that beyond 10\,pN there is little difference between the
force-extension curves when secondary structure is or is not allowed. This
secondary-structure induced low-force feature is also apparent in experimental
pulling curves for ssDNA at 500\,mM salt, but disappears at lower
salt.\cite{Huguet10} For comparison, fully complementary hairpins are
destabilized at about 16--17pN for the same salt conditions as
here.\cite{Huguet10} 

The area between the curve for ssDNA with intramolecular base-pairing forbidden
and the average curve with secondary structure present gives the average
stabilization of a 100-base strand due to secondary structure. At room
temperature, this stabilization is $0.23\,kT$ per base on average. Given this
value and the free-energy profiles in Fig.\ \ref{fig:FEL}, we can then estimate
the force at which the average free-energy change for unpeeling a base pair is
zero when secondary structure in the unpeeled strand is allowed. This gives a
correction of about 3\,pN at room temperature, although this may be a slight
underestimate, as we have not included the sequence that had the strongest
secondary structure in our estimate. 

Hence, our best estimate of the overstretching force 
for our model at 23$^\circ$C is $F_\mathrm{over}\approx 74$\,pN.
This value is satisfyingly close to the oft-quoted
65\,pN for room temperature overstretching, and much closer than has been
previously achieved for any coarse-grained model or atomistic simulation.  
More precisely, at 500\,mM salt, the concentration at which our coarse-grained 
model has been fitted, 
Wenner {\it et al.} obtain a value of 67\,pN,\cite{Wenner02} and 
and in the more recent study of Zhang {\it et al.} the onset and mid-point of 
the transitions were found to occur at 65\,pN and 68\,pN, 
respectively.\cite{Zhang12} 

\begin{figure} \includegraphics[width=8.4cm]{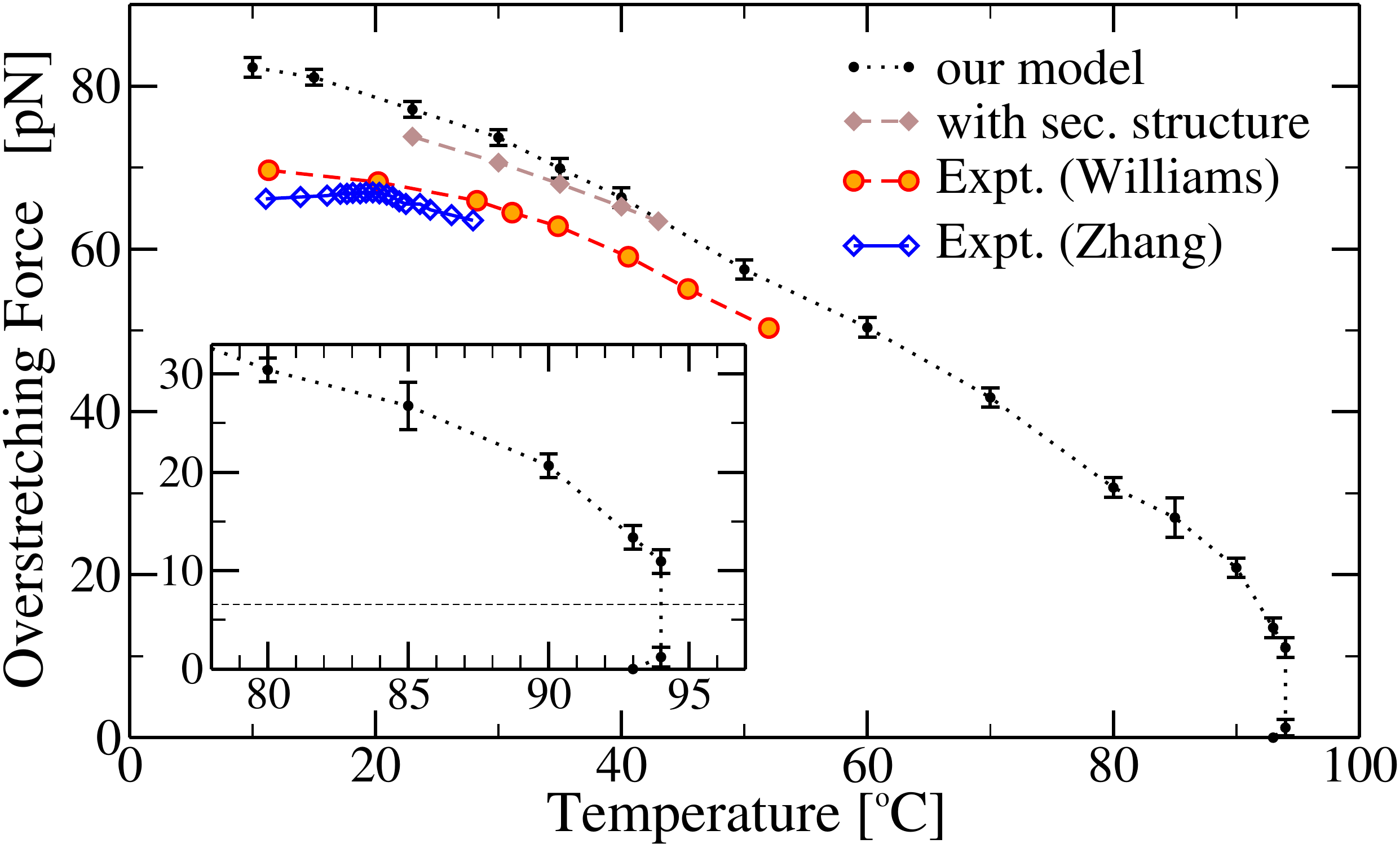}
\caption{\label{fig:Fover_T} Dependence of the overstretching force on
temperature. The main results for our model are from simulations where
intramolecular base-pairing was turned off, but we also include results where a
correction for secondary structure formation in the unpeeled chain has been
applied for temperatures betwen $23^\circ$C and $43^\circ$C --- above the
latter temperature there is no need for a correction as the secondary structure
is thermally unstable.  Also included are the experimental results from Refs.\
\onlinecite{Zhang12} and \onlinecite{Williams01} at a salt concentration of
500\,mM. Note that the results of Zhang {\it et al.} are for the onset (not the
midpoint) of the transition. The inset provides an expansion of the high
temperature region where non-monotonic behaviour is observed. The horizontal
line in the inset is the force at which ssDNA and dsDNA have the same
extension.
} \end{figure}

The same approach can be used to study the temperature dependence of
$F_\mathrm{over}$, including for room temperature and above the secondary
structure correction factor --- 
note that above $45^\circ$C the correction factor goes to zero, because by this
temperature all of the secondary structure in the single strand has melted.  
As the temperature is increased, the duplex is destabilized relative to ssDNA,
so $F_\mathrm{over}$ is expected to decrease. This behaviour is exactly what we
see for our model in Fig.\ \ref{fig:Fover_T}. Also included in this figure is
the experimental data of Refs.\ \onlinecite{Zhang12} and
\onlinecite{Williams01}.  Our results show a very similar temperature
dependence over the temperature range studied in experiments, albeit with our
results being displaced to slightly higher force by a similar amount to that
already seen at room temperature. In particular, Zhang {\it et al.} find a
slope of $-0.44$\,pN\,K$^{-1}$ at and above room temperature,\cite{Zhang12}
which is similar to the $-0.46$\,pN\,K$^{-1}$ that we obtain for our model.
Interestingly, Zhang {\it et al.} find a change in the sign of the slope below
20$^\circ$C which they interpret as a crossover to an S-DNA mechanism.  Our
results show no such change in slope, consistent with the fact that
overstretching always occurs by unpeeling in our model.

One particularly interesting feature at high temperature and low force is the
non-monotonic behaviour of $F_{\rm over}(T)$. There is a narrow range of
temperature just above the bulk (i.e.\ infinitely long strands) zero-force
melting temperature where there are two transitions as a function of force.
This behaviour is illustrated in the inset to Fig.\ \ref{fig:Fover_T} and also
in Fig.\ \ref{fig:FEL}(b) where there are two changes of sign in the slope of
the free-energy profile as a function of force for a temperature in this
window.  At low and high force it is favourable for the duplex to unpeel, but
there is an intermediate force range ($\sim$2.5--10\,pN at 94$^\circ$C) where
duplex DNA is most stable. This type of behaviour has been predicted for a
number of theoretical
models\cite{Rouzina01,Williams01,Hanke08,Rahi08,Einert10,Marenduzzo09,Marenduzzo10}
and occurs because which of dsDNA or ssDNA is more extensible depends on the
magnitude of force. At low force dsDNA is more extensible because it has a
larger persistence length and so has a smaller entropy cost for aligning with
the force. However, when the extension approaches the contour length of dsDNA
the force required to further extend the duplex grows rapidly, and ssDNA
becomes more extensible. The crossover between these two regimes occurs at
approximately 6.5\,pN (i.e.\ when the two force-extension curves in Fig.\
\ref{fig:Fvx_roomT}(a) cross). For forces below 6.5\,pN, dsDNA is stabilized
with respect to ssDNA by increasing force, but above this value of the force,
the opposite is true.  Hence, the turning point in $F_\mathrm{over}(T)$ is
expected to occur at this force.

\begin{figure}
\includegraphics[width=8.5cm]{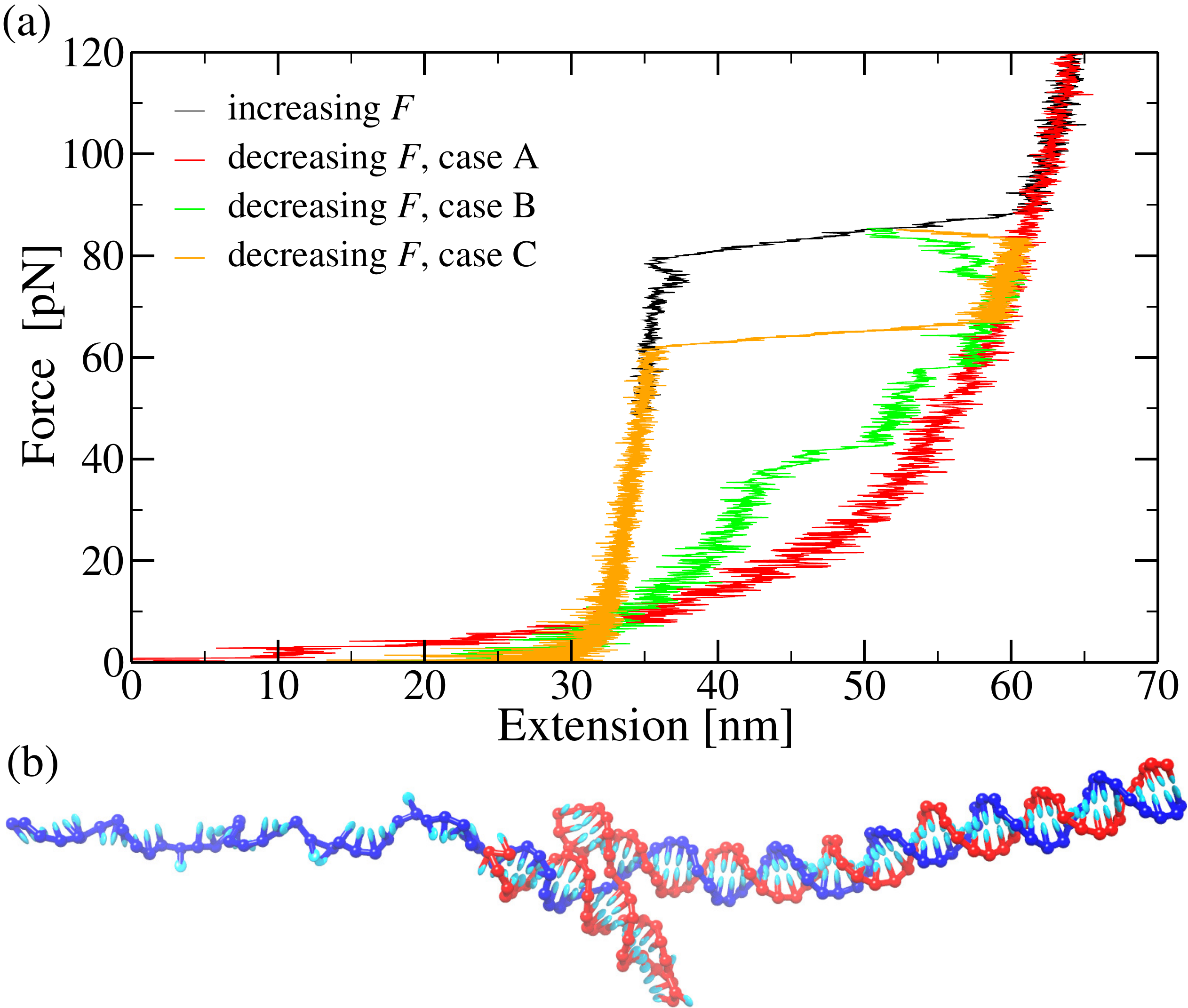}
\caption{\label{fig:reversible_roomT} 
(a) Force-extension curves illustrating the irreversibility in our model. 
In all the simulations the magnitude of the pulling rate is 
$2.43\times 10^{-6}$\,pN/time step. The increasing force run starts from a 
fully-bonded duplex, and provides the starting configurations for the decreasing
force simulations.
In case A, the force is decreased after reaching 146\,pN, whereas
in cases B and C the force is decreased after reaching 85.3\,pN. 
In case C intramolecular base pairing is not allowed, preventing secondary 
structure formation in the unpeeled chain.
(b) The final zero-force configuration for case B.
}
\end{figure}

One of the features in experiments when overstretching occurs by unpeeling is
that the transition is not fully reversible, and that on reducing the force
from above $F_\mathrm{over}$ the system does not trace out exactly the same
force-extension curve. One can envisage two possible 
sources for this lack of reversibility.  Firstly, if the
unpeeled strand has formed some secondary structure (e.g.\ hairpins), the
intramolecular base pairs involved would first have to be broken. 
Secondly, even for an unstructured unpeeled strand the reassociation dynamics
have the potential to be slow, particularly if 
the strands are long, as is typically the case in experiments. 

In Fig.\ \ref{fig:reversible_roomT}(a) we show
force-extension curves when the force is first increased and then decreased 
after reaching different stages in the overstretching process.
If the force is decreased
after overstretching when the non-force-bearing strand
has dissociated, the resulting force-extension curve just follows that for 
ssDNA (Case A), because reassociation of the separated chains does not occur.
Also depicted in Fig.\ \ref{fig:reversible_roomT}(a) are two cases when the
force decrease starts roughly halfway through the overstretching transition.
In both cases the extension, perhaps somewhat surprisingly, initially continues
to increase. As mentioned earlier, the dynamic pulling simulations
somewhat overshoot the overstretching transition. So,
even though the force is decreasing, initially it is still above the 
equilibrium overstretching force and so the molecule continues to unpeel. 
However, although the extension nearly reaches that for ssDNA, the chains do not
dissociate, because, as noted earlier, the last few base pairs are harder to 
melt. As the force decreases further, the two cases show quite different 
behaviour. In case B reassociation does occur but only partially and in 
``bursts''. Fig.\ \ref{fig:reversible_roomT}(b) depicts the final configuration
that results, clearly showing that full reassociation has been blocked by
secondary structure formation in the non-force-bearing strand. Similarly, the
bursts occur when sections involved in secondary structure open up, allowing 
further reassociation.

By contrast, in case C we prevent secondary structure in the unpeeled strand by
turning off intramolecular base pairing. Now when reassociation occurs, it does
so relatively quickly and goes to completion, albeit at a lower force than for
overstretching. Thus, for our model secondary structure formation is a much
more substantial contributor to irreversibility than the underlying dynamics of
reassociation. This finding is in agreement with Gross {\it et al.}\ who
found that prolonged stalls in the reannealing occurred at points at which
particularly stable secondary structure was possible in the unpeeled
strand.\cite{Gross11} However, we should note that our system is much smaller
than those typically studied in overstretching experiments (for example,
$\lambda$-DNA, which is often used in experiments, has 48\,502 base pairs).

\begin{figure}
\begin{center}
\includegraphics[width=8.5cm]{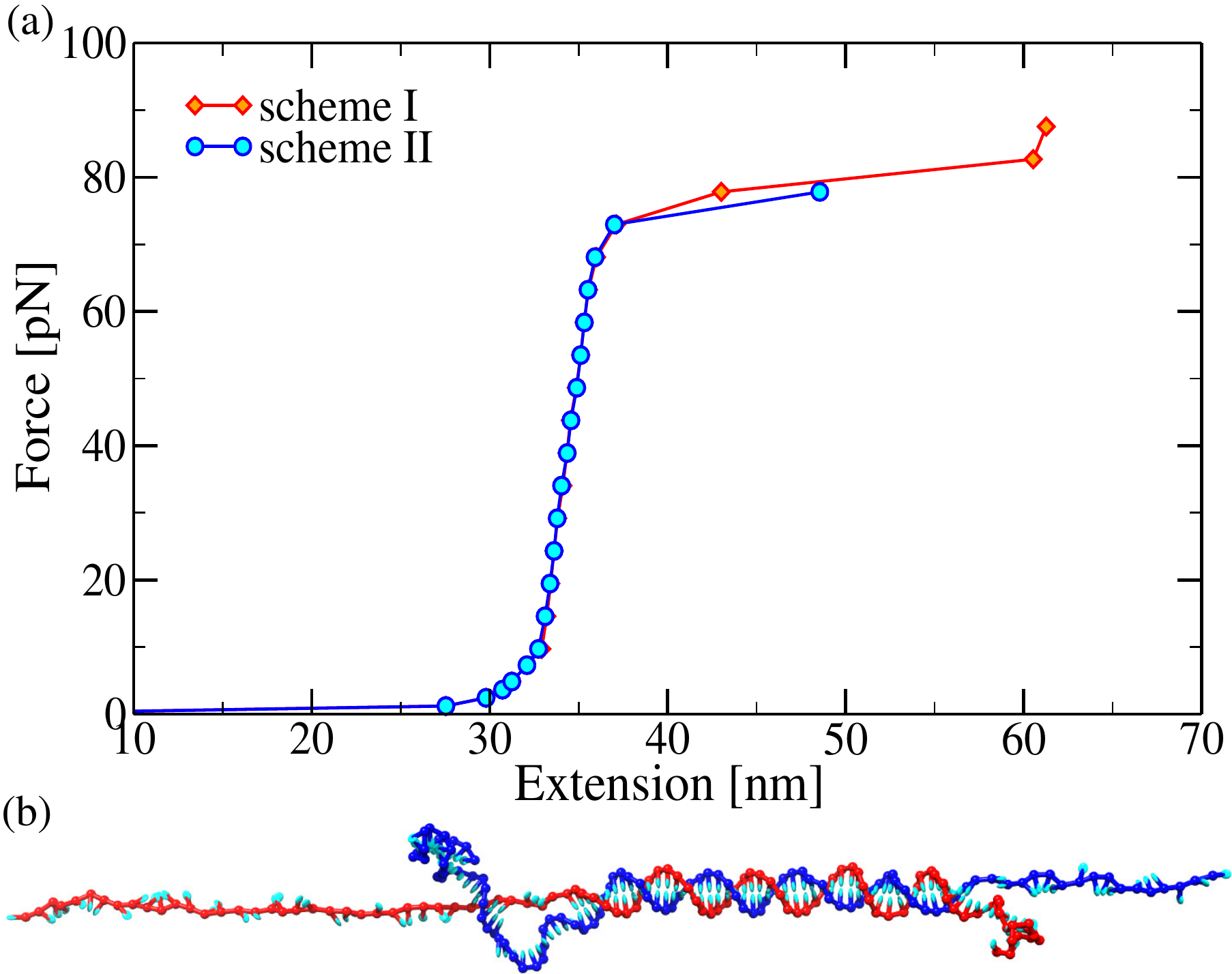}
\end{center}
\caption{\label{fig:Fx_schemeII} (a) Force-extension curve for pulling scheme II
obtained from Monte Carlos simulations compared to that for scheme I. 
There are no points above the midpoint of
overstretching because the system can no longer bear a force after
dissociation. (b) Snapshot near to the middle of the transition at $F=78$\,pN.
}
\end{figure}

Having considered pulling scheme I in considerable detail, we also examine the
other two pulling schemes in Fig.\ \ref{fig:schematics}. 
Firstly, for scheme II we expect a similar behaviour
to scheme I---there are still two free ends that allow unpeeling---except 
beyond the overstretching transition, where after dissociation of the two 
strands the system will no longer be able to bear a force. 
In Fig.\ \ref{fig:Fx_schemeII} we see that the overstretching transition occurs
at approximately the same force as for scheme I, and again occurs by unpeeling. 
Note that there are no points above the midpoint of overstretching due to 
this instability.

Scheme III is more interesting, because now overstretching can no longer occur
by unpeeling. Two remaining possible mechanisms for overstretching are
force-induced melting by bubble formation, and an S-DNA-like transition to an
overstretched but still associated duplex form. The former is expected to
require a larger force than unpeeling because both strands are still
force-bearing after bubble formation and so have less entropy than an unpeeled
strand. Interestingly, in room temperature experiments at 150\,mM NaCl 
where unpeeling cannot
occur and the system is not torsionally constrained, reversible overstretching 
still occurs at about 65\,pN.\cite{Paik11}

\begin{figure}
\includegraphics[width=8.5cm]{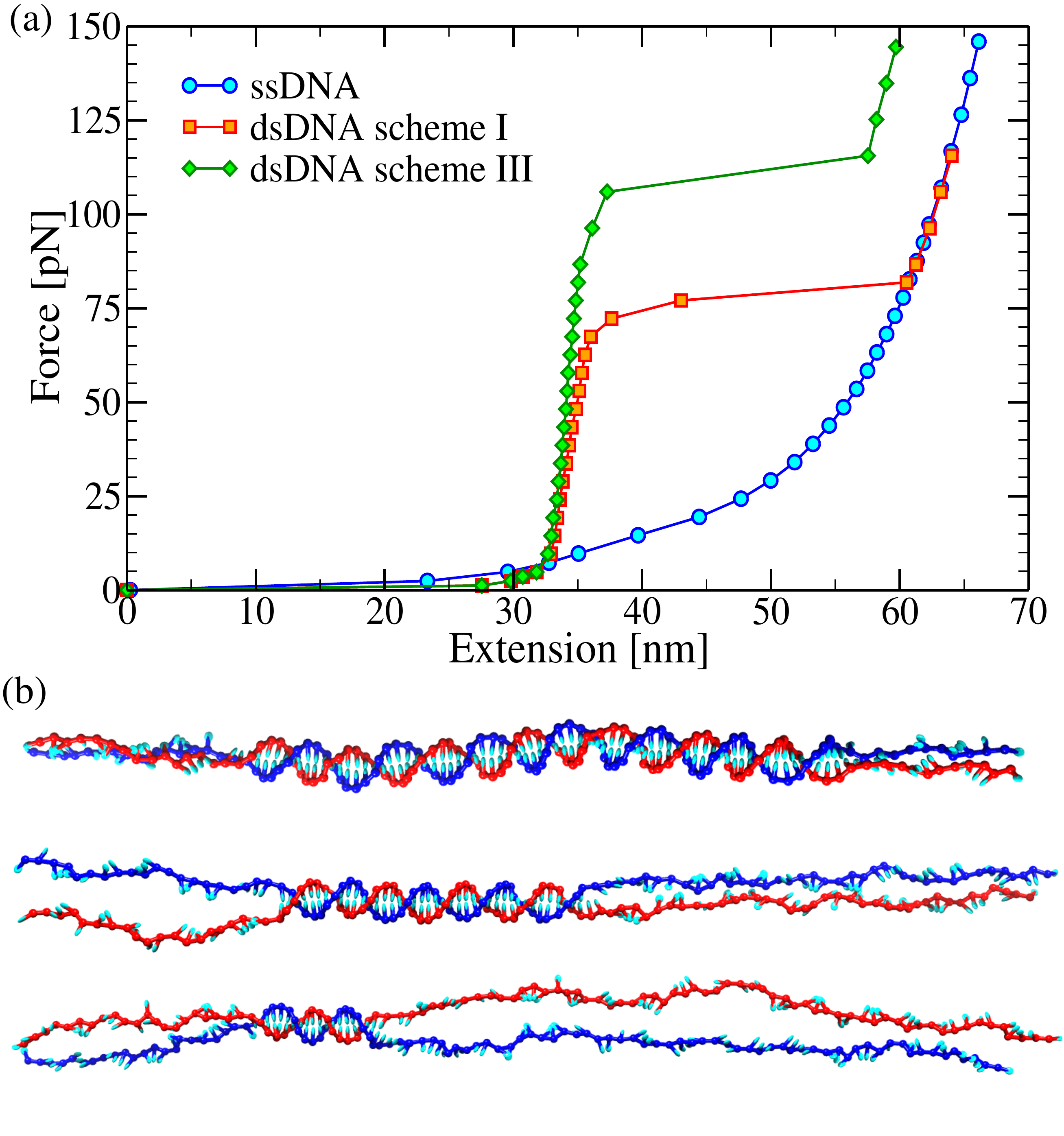}
\caption{\label{fig:Fx_schemeIII} 
(a) Force-extension curves and
(b) snapshots of the DNA configuration at $F=116$\,pN as it undergoes overstretching for 
pulling scheme III obtained from Monte Carlo simulations. 
For comparison, curves for pulling ssDNA and dsDNA in pulling 
scheme I have been added in (a).}
\end{figure}

Fig.\ \ref{fig:Fx_schemeIII}(a) shows the force-extension curve for our model
for pulling scheme III. The clear difference from scheme I and II is that
overstretching now occurs at significantly higher force, namely at about
110--115\,pN.  Although overstretching can no longer occur by unpeeling, Fig.\
\ref{fig:Fx_schemeIII}(b) shows that it is still by force-induced melting, but
in this case by bubble formation.  Notably, even though the duplex reaches
higher forces than in scheme I and II, there is still no sign of a transition
to an S-DNA-like overstretched form. Also, that the force-extension curve above
overstretching does not follow that for ssDNA is simply because each individual
strand only bears a force $F/2$ (Fig.\ \ref{fig:schematics}(c)),
and therefore the curve can be made to overlap with that for ssDNA by shifting it down
by a factor of 2.

\section{Conclusions}

In this paper we study the overstretching transition in a recently developed
coarse-grained model of DNA, which has previously been shown to give an
excellent description of the thermodynamic, structural and mechanical
properties of DNA.\cite{Ouldridge10,Ouldridge11} At room temperature 
and 500\,mM salt our model
undergoes overstretching by unpeeling at 74\,pN, just somewhat higher 
than the experimentally observed overstretching force, and much closer than the 
predictions for any other coarse-grained potential 
model\cite{Niewieczerzal09,Savin11b,Kocsis12,Hsu12}
or any atomistic simulations.\cite{,Harris05,Lohikoski05,Wereszczynski06,Luan08,Li09c,Santosh09,Rezac10,Balaeff11,Santosh11,Wolter11} We are also
able to reproduce the temperature dependence of the overstretching force well. 
This ability to accurately reproduce phenomena to which it was not directly 
fitted provides further validation of our model.

It is also important that we understand why our coarse-grained model does so
well in describing the overstretching transition. 
As emphasised in Section \ref{sect:thermo} to predict force-induced
melting correctly, one just needs to know the zero-force thermodynamics of 
melting and the force-extension curves of dsDNA and ssDNA. 
As our model describes all of the above reasonably well, its success 
in describing overstretching by force-induced melting is perhaps not surprising. 

The 6--7\,pN overestimation of the room temperature overstretching force
in our model is probably due to a couple of factors. Firstly, our description
of the ssDNA force-extension curve deviates slightly from fits to
experiment,\cite{Cocco04} 
in particular, underestimating the extension at larger forces. 
The net effect is to somewhat
underestimate the stabilization of ssDNA by force near to overstretching.
Secondly, the zero-force thermodynamics in our model has been designed to fit
the thermodynamic predictions of the SantaLucia nearest-neighbour
model,\cite{SantaLucia04} and so any errors in this model will be carried over
to our results. For example, Huguet {\it et al.} have shown that using the
SantaLucia model and fits to the force-extension curves of ssDNA and dsDNA
gives an approximately 1\,pN overestimation of DNA unzipping at room
temperature.\cite{Huguet10} As the difference in extension between the duplex
and ssDNA is less for the overstretching than for the unzipping geometry, one
would expect the error due to the use of the SantaLucia model to be larger for
overstretching, with the above result implying an overestimation of the order
of 4\,pN.  However, when Gross {\it et al.} applied a similar approach to model
their experimental overstretching force-extension curves, the model actually
underestimated the experimental results.\cite{Gross11}

Although in our simulations to locate the overstretching force, we had to turn
off secondary structure in the unpeeled strand in order to equilibrate the
system, we were able to estimate a correction to $F_\mathrm{over}$ from
calculations of the amount by which ssDNA is stabilized by secondary structure.
The correction factor for an average random sequence was 3\,pN at the solution
conditions for which our model has been fitted. However, for a sequence
specifically designed to form strong hairpins, the effect could be much larger,
and lead to a significant lowering of the overstretching force. Secondary
structure in the unpeeled strand also had a significant effect on the dynamics
with the associated free-energy barriers preventing the stretching simulations
reaching equilibrium, and leading to significant hysteresis on decreasing the
force. 

In all simulations where free ends are present, we see a force-induced melting
mechanism for overstretching occurring by unpeeling, one of the confirmed
experimental mechanisms.\cite{vanMameren09} However, no sign is ever seen of an
S-DNA-like transition to an overstretched duplex form.
This shortcoming leads to an overstretching force when there are no free ends
(i.e.\ scheme III) that is much higher than experiment, because the mechanism
is force-induced melting by bubble formation.  The reason for our model's
inability to exhibit an S-DNA-like state is harder to identify, as not much is
known about the detailed structure of S-DNA. One suggestion is that S-DNA might
adopt a ``ladder-like'' unstacked but base-paired
configuration.\cite{Thundat94,Smith96} However, if this were the case one might
expect our model to be able to reproduce this, as it describes the
thermodynamics of stacking and base-pairing well. Furthermore, the reason why
such a form is not stable in our model is simply that stacking provides a
larger free energetic contribution to the stability of B-DNA than base pairing
and that the overstretched unpeeled state retains a significant fraction of its
stacking (Fig.\ \ref{fig:V_roomT}). This stacking occurs both in the
non-force-bearing unpeeled strand that is totally free to stack and in the
force-bearing strand where short runs of stacked bases can align their
backbones along the force axis. Even when there are no free ends, for which
force-induced melting in our model occurs at higher force by bubble formation
with less stacking retained because both strands are force-bearing, an
unstacked but base-paired duplex is still not observed. Our results strongly
suggest that S-DNA does not correspond to such an unstacked but base-paired
configuration, but instead probably to a more exotic structure.  A plausible
candidate structure for S-DNA might be ``zip-DNA'',\cite{Lohikoski05,Balaeff11}
where the chains are no longer base-paired but interdigitate allowing stacking
between bases on different strands. Our model does not include the possibility
of this more unusual stacking interaction, and so of course would not be able
to reproduce any such behaviour.

\begin{acknowledgments}
The authors are grateful for financial support from the EPSRC and University
College, and for helpful discussions with Felix Ritort. 
\end{acknowledgments}


\begin{thebibliography}{10}
\bibitem{nocomplement}%
  \BibitemOpen
  \bibinfo {note} {This could also be achieved by choosing a sequence with no
  complementary bases in the same strand.}\BibitemShut{Stop}%
\bibitem{repetitive}%
  \BibitemOpen
  \bibinfo {note} {In some simulations with repetitive sequences (not reported
  here) we have seen a tendency for the unpeeled strand to rebind to the
  force-bearing strand in a non-native site for short stretches (short so that
  the helix of the force-bearing strand can still approximately allign along
  the direction of force).}\BibitemShut{Stop}%
\end{thebibliography}
%

\end{document}